\input harvmac.tex
%\draftmode
\lref\MotlTH{
L.~Motl,
``Proposals on nonperturbative superstring interactions,''
arXiv:hep-th/
9701025.
%%CITATION = HEP-TH 9701025;%%
S.~Sethi and L.~Susskind,
``Rotational invariance in the M(atrix) formulation of type IIB
theory,''
Phys.\ Lett.\ B {\bf 400}, 265 (1997)
[arXiv:hep-th/9702101].
%%CITATION = HEP-TH 9702101;%%
T.~Banks and N.~Seiberg,
``Strings from matrices,''
Nucl.\ Phys.\ B {\bf 497}, 41 (1997)
[arXiv:hep-th/9702187].
%%CITATION = HEP-TH 9702187;%%
}
\lref\TaylorIK{
W.~I.~Taylor,
``D-brane field theory on compact spaces,''
Phys.\ Lett.\ B {\bf 394}, 283 (1997)
[arXiv:hep-th/9611042].
%%CITATION = HEP-TH 9611042;%%
}
\lref\BanksVH{
T.~Banks, W.~Fischler, S.~H.~Shenker and L.~Susskind,
``M theory as a matrix model: A conjecture,''
Phys.\ Rev.\ D {\bf 55}, 5112 (1997)
[arXiv:hep-th/9610043].
%%CITATION = HEP-TH 9610043;%%
}
\lref\GreensiteGV{
J.~Greensite and F.~R.~Klinkhamer,
``New Interactions For Superstrings,''
Nucl.\ Phys.\ B {\bf 281}, 269 (1987).
%%CITATION = NUPHA,B281,269;%%
J.~Greensite and F.~R.~Klinkhamer,
``Contact Interactions In Closed Superstring Field Theory,''
Nucl.\ Phys.\ B {\bf 291}, 557 (1987).
%%CITATION = NUPHA,B291,557;%%
M.~B.~Green and N.~Seiberg,
``Contact Interactions In Superstring Theory,''
Nucl.\ Phys.\ B {\bf 299}, 559 (1988).
%%CITATION = NUPHA,B299,559;%%
J.~Greensite and F.~R.~Klinkhamer,
``Superstring Amplitudes And Contact Interactions,''
Nucl.\ Phys.\ B {\bf 304}, 108 (1988).
%%CITATION = NUPHA,B304,108;%%
}
\lref\GreenTC{
M.~B.~Green and J.~H.~Schwarz,
``Superstring Interactions,''
Nucl.\ Phys.\ B {\bf 218}, 43 (1983).
%%CITATION = NUPHA,B218,43;%%
M.~B.~Green, J.~H.~Schwarz and L.~Brink,
``Superfield Theory Of Type Ii Superstrings,''
Nucl.\ Phys.\ B {\bf 219}, 437 (1983).
%%CITATION = NUPHA,B219,437;%%
}
\lref\DiFrancescoNK{
P.~Di Francesco, P.~Mathieu and D.~Senechal,
``Conformal Field Theory.''
}
\lref\DixonQV{
L.~J.~Dixon, D.~Friedan, E.~J.~Martinec and S.~H.~Shenker,
``The Conformal Field Theory Of Orbifolds,''
Nucl.\ Phys.\ B {\bf 282}, 13 (1987).
%%CITATION = NUPHA,B282,13;%%
}
\lref\BershadskyFV{
M.~Bershadsky and A.~Radul,
``Conformal Field Theories With Additional Z(N) Symmetry,''
Int.\ J.\ Mod.\ Phys.\ A {\bf 2}, 165 (1987).
%%CITATION = IMPAE,A2,165;%%
}
\lref\MartinecCD{
E.~J.~Martinec,
``String Calculus: Conformal Field Theory As A Tool In String Theory,''
EFI-87-45-CHICAGO.
}
\lref\DijkgraafVV{
R.~Dijkgraaf, E.~Verlinde and H.~Verlinde,
``Matrix string theory,''
Nucl.\ Phys.\ B {\bf 500}, 43 (1997)
[arXiv:hep-th/9703030].
%%CITATION = HEP-TH 9703030;%%
}
\lref\DijkgraafNW{
R.~Dijkgraaf and L.~Motl,
``Matrix string theory, contact terms, and superstring field theory,''
arXiv:hep-th/0309238.
%%CITATION = HEP-TH 0309238;%%
}
\lref\GrossIA{
D.~J.~Gross and A.~Jevicki,
``Operator Formulation Of Interacting String Field Theory,''
Nucl.\ Phys.\ B {\bf 283}, 1 (1987),
%%CITATION = NUPHA,B283,1;%%
D.~J.~Gross and A.~Jevicki,
``Operator Formulation Of Interacting String Field Theory. 2,''
Nucl.\ Phys.\ B {\bf 287}, 225 (1987).
%%CITATION = NUPHA,B287,225;%%
}
\lref\LeClairSP{
A.~LeClair, M.~E.~Peskin and C.~R.~Preitschopf,
``String Field Theory On The Conformal Plane. 1. Kinematical Principles,''
Nucl.\ Phys.\ B {\bf 317}, 411 (1989),
%%CITATION = NUPHA,B317,411;%%
A.~LeClair, M.~E.~Peskin and C.~R.~Preitschopf,
``String Field Theory On The Conformal Plane. 2. Generalized Gluing,''
Nucl.\ Phys.\ B {\bf 317}, 464 (1989).
%%CITATION = NUPHA,B317,464;%%
}
\lref\WittenCC{
E.~Witten,
``Noncommutative Geometry And String Field Theory,''
Nucl.\ Phys.\ B {\bf 268}, 253 (1986).
%%CITATION = NUPHA,B268,253;%%
}
\lref\PankiewiczPG{
A.~Pankiewicz,
``Strings in plane wave backgrounds,''
Fortsch.\ Phys.\  {\bf 51}, 1139 (2003)
[arXiv:hep-th/0307027].
%%CITATION = HEP-TH 0307027;%%
J.~C.~Plefka,
``Lectures on the plane-wave string / gauge theory duality,''
Fortsch.\ Phys.\  {\bf 52}, 264 (2004)
[arXiv:hep-th/0307101].
%%CITATION = HEP-TH 0307101;%%
J.~M.~Maldacena,
``TASI 2003 lectures on AdS/CFT,''
arXiv:hep-th/0309246.
%%CITATION = HEP-TH 0309246;%%
M.~Spradlin and A.~Volovich,
``Light-cone string field theory in a plane wave,''
arXiv:hep-th/0310033.
%%CITATION = HEP-TH 0310033;%%
D.~Sadri and M.~M.~Sheikh-Jabbari,
``The plane-wave / super Yang-Mills duality,''
arXiv:hep-th/0310119.
%%CITATION = HEP-TH 0310119;%%
M.~x.~Huang,
``Aspects of supersymmetric gauge theory and string theory,''
arXiv:hep-th/0404211.
%%CITATION = HEP-TH 0404211;%%
}
\lref\RoibanXR{
R.~Roiban, M.~Spradlin and A.~Volovich,
``On light-cone SFT contact terms in a plane wave,''
JHEP {\bf 0310}, 055 (2003)
[arXiv:hep-th/0211220].
%%CITATION = HEP-TH 0211220;%%
}
\lref\GomisKJ{
J.~Gomis, S.~Moriyama and J.~w.~Park,
``SYM description of pp-wave string interactions: Singlet sector and
arbitrary impurities,''
Nucl.\ Phys.\ B {\bf 665}, 49 (2003)
[arXiv:hep-th/0301250].
%%CITATION = HEP-TH 0301250;%%
}
\lref\HorowitzYZ{
G.~T.~Horowitz and A.~Strominger,
``Translations As Inner Derivations And Associativity Anomalies In
Open String Field Theory,''
Phys.\ Lett.\ B {\bf 185}, 45 (1987).
%%CITATION = PHLTA,B185,45;%%
}
\lref\HataWA{
H.~Hata and S.~Moriyama,
``Observables as twist anomaly in vacuum string field theory,''
JHEP {\bf 0201}, 042 (2002)
[arXiv:hep-th/0111034].
%%CITATION = HEP-TH 0111034;%%
H.~Hata, S.~Moriyama and S.~Teraguchi,
``Exact results on twist anomaly,''
JHEP {\bf 0202}, 036 (2002)
[arXiv:hep-th/0201177].
%%CITATION = HEP-TH 0201177;%%
}
\lref\ArutyunovGT{
G.~E.~Arutyunov and S.~A.~Frolov,
``Virasoro amplitude from the S(N) R**24 orbifold sigma model,''
Theor.\ Math.\ Phys.\  {\bf 114}, 43 (1998)
[arXiv:hep-th/9708129].
%%CITATION = HEP-TH 9708129;%%
G.~E.~Arutyunov and S.~A.~Frolov,
``Four graviton scattering amplitude from S(N) R**8 supersymmetric
orbifold sigma model,''
Nucl.\ Phys.\ B {\bf 524}, 159 (1998)
[arXiv:hep-th/9712061].
%%CITATION = HEP-TH 9712061;%%
}
\lref\GeorgiouTY{
G.~Georgiou and G.~Travaglini,
``Fermion BMN operators, the dilatation operator of N=4 SYM, and
pp-wave string interactions,''
arXiv:hep-th/0403188.
%%CITATION = HEP-TH 0403188;%%
}
\lref\GrossMH{
D.~J.~Gross, A.~Mikhailov and R.~Roiban,
``A calculation of the plane wave string Hamiltonian from N = 4
super-Yang-Mills theory,''
JHEP {\bf 0305}, 025 (2003)
[arXiv:hep-th/0208231].
%%CITATION = HEP-TH 0208231;%%
J.~Gomis, S.~Moriyama and J.~w.~Park,
``SYM description of SFT Hamiltonian in a pp-wave background,''
Nucl.\ Phys.\ B {\bf 659}, 179 (2003)
[arXiv:hep-th/0210153].
%%CITATION = HEP-TH 0210153;%%
J.~Pearson, M.~Spradlin, D.~Vaman, H.~Verlinde and A.~Volovich,
``Tracing the string: BMN correspondence at finite J**2/N,''
JHEP {\bf 0305}, 022 (2003)
[arXiv:hep-th/0210102].
%%CITATION = HEP-TH 0210102;%%
}

\Title{\vbox{\baselineskip12pt
\hbox{CALT-68-2498}\hbox{hep-th/0405091}}}
{\vbox{\centerline{Comments on Supersymmetry Algebra}\vskip6pt
\centerline{and Contact Term in Matrix String Theory}}}

\centerline{Sanefumi Moriyama\footnote{$^\dagger$}
{moriyama@theory.caltech.edu}}
\bigskip\centerline{\it California Institute of Technology 452-48,
Pasadena, CA 91125}
\vskip .3in

\centerline{Abstract}
Following hep-th/0309238 relating the matrix string theory to the
light-cone superstring field theory, we write down two supercharges in
the matrix string theory explicitly. After checking the supersymmetry
algebra at the leading order, we proceed to discuss higher-order
contact terms.

\Date{5/2004}
\newsec{Introduction}
String field theory is a standard off-shell formalism to describe
fundamental string interactions in the perturbative way.
Recently its importance is recognized in various contexts.
Among others, the success in the PP-wave/SYM correspondence is
remarkable. (See \PankiewiczPG\ for reviews and \GeorgiouTY\ for
recent progress.)
The dilatation matrix elements calculated in perturbative SYM theory
were found to match completely with the Hamiltonian matrix elements of
the corresponding states in PP-wave light-cone string field theory
after a basis change \GrossMH.
What is surprising is that, the matching goes beyond the cubic
interaction: even the contact term was found to match if we truncate
the intermediate states to a particular sector \RoibanXR\GomisKJ.
However, we have to confess that our understanding of the contact
terms is very restricted and the meaning of this truncation still
needs to be clarified.
Although there is a canonical method to discuss the contact terms
using the supersymmetry algebra, various questions related to its
general structure remain unanswered.
The main difficulty consists in its notorious complexity of the
algebra.

Last summer, Dijkgraaf and Motl \DijkgraafNW\ revisited the matrix
string theory \MotlTH\DijkgraafVV, another proposal to describe the
fundamental string interaction.
They related very extensively the matrix string theory to the
light-cone superstring field theory in the flat space \GreenTC\ and
also came up with an interesting proposal on the contact term.
Since on the string field theory side we have a systematic method to
discuss the contact term from the supersymmetry algebra, it is very
tantalizing to apply this method directly to the matrix string
theory.
This is our main theme in this paper.
We hope we can avoid the complexity of the algebra on the string field
theory side in this way and present a systematic calculation to prove
or disprove the contact term proposal of \DijkgraafNW.
We will see later that our method, though still has many problems,
looks somewhat promising.

Another motivation for our present work is as follows.
In Witten's bosonic cubic string field theory \WittenCC, besides the 
operator formalism \GrossIA, we have another CFT formalism
\LeClairSP.
It should be good to have more than one description, because physical
observables sometimes appear to be anomalous in the string field theory
\HorowitzYZ\HataWA.
We hope the matrix string theory will serve as another formalism for
the light-cone superstring field theory and studying the matrix string
theory will give us insights on the light-cone string field theory,
and vice versa.

The contents of the rest of this paper is as follows.
We start with reviewing the light-cone superstring field theory in
section two.
Instead of going into the details of this theory, we only focus on the
conceptual part.
Especially, we explain how light-cone superstring interactions are
determined from supersymmetry algebra.
After that, we proceed to the matrix string theory in section three,
explaining our strategy to the interaction terms by mimicking the
light-cone superstring field theory.
Following this strategy, we check the supersymmetry algebra at the
first non-trivial stringy level in section four and proceed to discuss
some problems in calculating the contact term in section five.

\newsec{Conceptual review of light-cone superstring field theory}
In this section we shall review the light-cone superstring field
theory \GreenTC.
The light-cone superstring field theory is constructed by generalizing
the light-cone quantization of string theory in Green-Schwarz
formalism.
Hence, the start point is the covariant string worldsheet action with
spacetime supersymmetries.
Since we cannot quantize the action covariantly, we have to choose the
light-cone gauge.
After the gauge-fixing, the light-cone Hamiltonian and two
supercharges read
\eqn\zeroth{\eqalign{
&H_0=\int_0^{2\pi}{d\sigma\over 2\pi}
{1\over 4}\Bigl((P^i)^2+(X^{i\prime})^2
+\theta^{\alpha}\theta^{\alpha\prime}
+\bar\theta^{\alpha}\bar\theta^{\alpha\prime}\Bigr),\cr
&Q^{\dot\alpha}_0=\int_0^{2\pi}{d\sigma\over 2\pi}
\gamma^i_{\alpha\dot\alpha}\theta^{\alpha}(P^i+X^{i\prime}),\cr
&\tilde Q^{\dot\alpha}_0=\int_0^{2\pi}{d\sigma\over 2\pi}
\gamma^i_{\alpha\dot\alpha}\bar\theta^{\alpha}(P^i-X^{i\prime}),}}
with $\theta^\alpha$ being the Green-Schwarz fermions.
These charges satisfy the supersymmetry algebra
\eqn\zeroalg{
\{Q^{\dot\alpha}_0,Q^{\dot\beta}_0\}
=\{\tilde Q^{\dot\alpha}_0,\tilde Q^{\dot\beta}_0\}
=2H\delta^{\dot\alpha\dot\beta},\qquad
[Q^{\dot\alpha}_0,H_0]=[\tilde Q^{\dot\alpha}_0,H_0]=0,}
with two supercharges being independent:
\eqn\zeroindep{\{Q^{\dot\alpha}_0,\tilde Q^{\dot\beta}_0\}=0.}
Note that in the closed string theory we have the level-matching
condition, relating the left-moving modes to right-moving modes.
The algebra holds up to this constraint.

So far we only have the free part.
To construct the light-cone superstring field theory, we have to
include the interaction part
\eqn\full{\eqalign{
H&=H_0+g_sH_1+g_s^2H_2+\cdots,\cr
Q^{\dot\alpha}&=Q^{\dot\alpha}_0+g_sQ^{\dot\alpha}_1
+g_s^2Q^{\dot\alpha}_2+\cdots,\cr
\tilde Q^{\dot\alpha}&=\tilde Q^{\dot\alpha}_0
+g_s\tilde Q^{\dot\alpha}_1
+g_s^2\tilde Q^{\dot\alpha}_2+\cdots,}}
respecting the full supersymmetry algebra order by order
\eqn\fullalg{
\{Q^{\dot\alpha},Q^{\dot\beta}\}
=\{\tilde Q^{\dot\alpha},\tilde Q^{\dot\beta}\}
=2H\delta^{\dot\alpha\dot\beta},\qquad
[Q^{\dot\alpha},H]=[\tilde Q^{\dot\alpha},H]=0.}
Besides, as in the free case, we require two supercharges to be
independent:
\eqn\twoindepend{\{Q^{\dot\alpha},\tilde Q^{\dot\beta}\}=0.}
More explicitly, if we concentrate only on one of the two
supercharges, \fullalg\ will become
\eqn\susyexpand{\eqalign{
\{Q^{\dot\alpha}_0,Q^{\dot\beta}_0\}
&=2H_0\delta^{\dot\alpha\dot\beta},\qquad
[Q^{\dot\alpha}_0,H_0]=0,\cr
\{Q^{\dot\alpha}_0,Q^{\dot\beta}_1\}
+\{Q^{\dot\alpha}_1,Q^{\dot\beta}_0\}
&=2H_1\delta^{\dot a\dot b},\qquad
[Q^{\dot\alpha}_0,H_1]+[Q^{\dot\alpha}_1,H_0]=0,\cr
\{Q^{\dot\alpha}_0,Q^{\dot\beta}_2\}
+\{Q^{\dot\alpha}_1,Q^{\dot\beta}_1\}
+\{Q^{\dot\alpha}_2,Q^{\dot\beta}_0\}
&=2H_2\delta^{\dot\alpha\dot\beta},\qquad
[Q^{\dot\alpha}_0,H_2]+[Q^{\dot\alpha}_1,H_1]
+[Q^{\dot\alpha}_2,H_0]=0.}}

At the first non-trivial order, $Q^{\dot\alpha}_1$,
$\tilde Q^{\dot\alpha}_1$ and $H_1$ were solved in \GreenTC.
Since these interaction vertices relate two strings Hilbert space to
one string Hilbert space, it is best to represent them as ket states:
\eqn\firstSFT{\eqalign{
|H_1\rangle_{123}
&=\bar Z^iZ^jv^{ij}(\Lambda)|V\rangle_{123},\cr
|Q^{\dot\alpha}_1\rangle_{123}
&=\bar Z^is^{i\dot\alpha}(\Lambda)|V\rangle_{123},\cr
|\tilde Q^{\dot\alpha}_1\rangle_{123}
&=Z^i\tilde s^{i\dot\alpha}(\Lambda)|V\rangle_{123}.}}
Here $|V\rangle_{123}$ represents the geometrical overlapping of two
strings and one string.
The quantities $Z^i$, $\bar Z^i$ and $\Lambda^a$ are, roughly
speaking, the renormalized operators of holomorphic part of bosonic
momentum $P^i-X^{i\prime}$, anti-holomorphic part $P^i+X^{i\prime}$
and fermionic momentum $\lambda$ evaluated at the interaction point
respectively and $v^{ij}$, $s^{i\dot\alpha}$ and
$\tilde s^{i\dot\alpha}$ are some complicate polynomials of
$\Lambda$.

Apparently, there are many questions related to this construction
\GreensiteGV.
Does this expansion end at this order?
If not, how do the terms of the next order $Q^{\dot\alpha}_2$,
$\tilde Q^{\dot\alpha}_2$ and $H_2$ look like?
Can we compute the terms up to arbitrary order or do we have to
calculate order by order?
However, the present formalism is complicated and we do not have a
satisfactory answer to these questions.

\newsec{Matrix string theory}
Let us turn to the matrix string theory \DijkgraafVV.
The matrix string theory proposal was motivated by compactifying
\TaylorIK\MotlTH\ M(atrix) theory \BanksVH\ on one dimension.
The definition of the matrix string theory is the low energy effective
action of the two-dimensional super Yang-Mills theory.
When the theory flows to the IR limit, it will look like a free theory
of string bits propagating in the moduli space with the Hamiltonian
and the supercharges given by
\eqn\mszeroth{\eqalign{
H_0&=\sum_{m=1}^N\int_0^{2\pi}{d\sigma\over 2\pi}
{1\over 4}\Bigl((P^i_m)^2+(X^{i\prime}_m)^2
+\theta^\alpha_m\theta^{\alpha\prime}_m
+\bar\theta^\alpha_m\bar\theta^{\alpha\prime}_m\Bigr),\cr
Q^{\dot\alpha}_0&=\sum_{m=1}^N\int_0^{2\pi}{d\sigma\over 2\pi}
\gamma^i_{\alpha\dot\alpha}\theta^\alpha_m(P^i_m+X^{i\prime}_m),\cr
\tilde Q^{\dot\alpha}_0&=\sum_{m=1}^N\int_0^{2\pi}{d\sigma\over 2\pi}
\gamma^i_{\alpha\dot\alpha}\bar\theta^\alpha_m(P^i_m-X^{i\prime}_m).}}
Here the index $m$ denotes each of the string bits.

To incorporate the interaction terms, we have to return slightly from
the complete IR limit and consider the least irrelevant operator.
The first non-trivial order of the Hamiltonian in matrix string theory
is conjectured to be
\eqn\Hone{H_1=\sum_{m<n}\int_0^{2\pi}{d\sigma\over 2\pi}
([\tau^i\bar\tau^j]\Sigma^i\bar\Sigma^j)_{(mn)}.}
Here all the fields are twist fields or spin fields exchanging $m$-th
and $n$-th string bits.
If we take $[\sigma\bar\sigma](z,\bar z)$ to be the twist field in the
bosonic sector, which corresponds to the anti-periodic ground state
$|0\rangle_{\rm AP}$, $[\tau^i\bar\tau^j](z,\bar z)$ is the excited
twist field, corresponding to
$\alpha^i_{-1/2}\bar\alpha^j_{-1/2}|0\rangle_{\rm AP}$.
Note that the twist field $[\sigma\bar\sigma](z,\bar z)$ do not
decouple to the holomorphic part $\sigma(z)$ and the anti-holomorphic
part $\bar\sigma(\bar z)$, though we symbolically use this notation.
The spin fields of the fermionic sector can be classified by 
${\bf 8_v}$ and ${\bf 8_c}$, therefore we denote them as $\Sigma^i$
and $\Sigma^{\dot\alpha}$.
Here we take fields with integer (half-integer) dimension to be bosons
(fermions).

One evidence for this interaction term is the dimension counting.
The conformal dimension of the least irrelevant operator \Hone\ is
(3/2,3/2).
This fact corresponds beautifully to the fact that the first
non-trivial interaction should be proportional to the string coupling
$g_s$.
Another evidence is the idea of supersymmetry.
%\eqn\susy{[Q^{\dot\alpha}_0,\sigma\Sigma^{\dot\beta}]
%+[\sigma\Sigma^{\dot\alpha},Q_0^{\dot\beta}]
%=\tau^i\Sigma^i\delta^{\dot\alpha\dot\beta},\qquad
%\{Q^{\dot\alpha}_0,\tau^i\Sigma^i\}
%=\partial(\sigma\Sigma^{\dot\alpha}).}
Also, it was shown \ArutyunovGT\ that the Veneziano amplitude is
reproduced from this interaction term.

Besides, in \DijkgraafNW\ the relation between the matrix string
theory and the light-cone superstring field theory was made explicit.
They found the quantities $v^{ij}$, $s^{i\dot\alpha}$ and
$\tilde s^{i\dot\alpha}$ appearing in the first non-trivial order in
the light-cone superstring field theory \firstSFT\ are related
respectively to
\eqn\relation{v^{ij}\Leftrightarrow\Sigma^i\bar\Sigma^j,\qquad
s^{i\dot\alpha}\Leftrightarrow\Sigma^{\dot\alpha}\bar\Sigma^i,
\qquad\tilde s^{i\dot\alpha}
\Leftrightarrow\Sigma^i\bar\Sigma^{\dot\alpha},}
in the matrix string theory.
Therefore, comparing \Hone\ with the first line of \firstSFT, we find
a complete match.
Moreover, it is very plausible to write down the supercharges at the
first non-trivial order as
\eqn\Qone{\eqalign{
Q_1^{\dot\alpha}&=-\sqrt{2}
\sum_{m<n}\int_0^{2\pi}{d\sigma\over 2\pi}
([\sigma\bar\tau^i]\Sigma^{\dot\alpha}\bar\Sigma^i)_{(mn)},\cr
\tilde Q_1^{\dot\alpha}&=-\sqrt{2}
\sum_{m<n}\int_0^{2\pi}{d\sigma\over 2\pi}
([\tau^i\bar\sigma]\Sigma^i\bar\Sigma^{\dot\alpha})_{(mn)}.}}
where $[\tau^i\bar\sigma](z,\bar z)$ and
$[\sigma\bar\tau^i](z,\bar z)$ correspond respectively to
$\alpha^i_{-1/2}|0\rangle_{\rm AP}$ and
$\bar\alpha^i_{-1/2}|0\rangle_{\rm AP}$.

Note that, we can rewrite the integral of $\sigma$ in terms of the
holomorphic coordinate $z$ or the anti-holomorphic one $\bar z$.
Although for a conserved charge
\eqn\charge{Q=\int_C dzj+d\bar z\bar j,}
$j$ and $\bar j$ must satisfy
\eqn\conserved{\bar\partial j-\partial\bar j=0,}
we have not succeeded in separating the expressions of \Hone\ or
\Qone\ into $j$ and $\bar j$.
So, in \Hone\ and \Qone\ the integral contour is considered to be
strictly the circle centered at the origin.

It is very important to work out the supersymmetry algebra, because
the supersymmetry algebra is the main guiding principle in
constructing the light-cone superstring field theory.
As reviewed in the previous section, to investigate the contact terms,
we have to calculate \susyexpand\ in higher orders.
For that purpose we start with the leading order.
In the following sections, we shall elaborate the supersymmetry
algebra at the first non-trivial order and then proceed to the contact
terms.

\newsec{Supersymmetry at the leading order}
In this section, we shall check the supersymmetry algebra at the
leading order.
Note that, conceptually we have already known the result, because at
this order the supersymmetry algebra was worked out in the light-cone
superstring field theory by \GreenTC, and its relation to the matrix
string theory was explored by \DijkgraafNW.
Technically, the supersymmetry algebra was mostly worked out in the
original paper \DijkgraafVV, except a few details about including both
the holomorphic and anti-holomorphic parts and about confirming the
independence between two supercharges \twoindepend.
However, we believe that it is important to make the leading order
supersymmetry algebra clear at this point.

Let us begin with the free part.
We would like to show the first line of \susyexpand.
First of all, we have to rewrite the expressions \mszeroth\ in terms
of the holomorphic coordinates:
\eqn\zerohol{\eqalign{
&H_0=-{1\over 2}\int{dz\over 2\pi}z\Bigl((\partial X^i)^2
+\theta^\alpha\partial\theta^\alpha\Bigr),\cr
&Q_0^{\dot\alpha}=\int{dz\over 2\pi}
\sqrt{iz}\gamma^i_{\alpha\dot\alpha}\theta^\alpha\partial X^i.}}
Here we drop the indices for the string bit ($m$ in \mszeroth).
Since there is no interaction between string bits in the free case, we
pick up a particular string bit and sum them up after the
calculation.
Also, the Hamiltonian can be written in terms of either the
holomorphic variable or the anti-holomorphic one because we always
restrict our Hilbert space to the subspace satisfying the
level-matching condition, but we shall choose the holomorphic one
here.

In the calculation we need the OPE of $\partial X^i$ and $\theta^a$
between themselves.
Since they are free fields, the OPE are just
\eqn\free{\eqalign{
\partial X^i(z)\cdot\partial X^j(0)
&\sim-{\delta^{ij}\over z^2}+
:\partial X^i\partial X^j:(0)+O(z),\cr
\theta^\alpha(z)\cdot\theta^\beta(0)
&\sim{\delta^{\alpha\beta}\over z}+:\theta^\alpha\theta^\beta:(0)
+O(z).}}
Using these OPE we can calculate the commutation relation by the usual
trick of deforming the contours at the same radius into a small one
encircling the singularity.
The relation
$\{Q_0^{\dot\alpha},Q_0^{\dot\beta}\}
=2H_0\delta^{\dot\alpha\dot\beta}$ 
is fairly easy to show, though in calculating $[Q_0^{\dot\alpha},H_0]$
we find
\eqn\zeroexample{[Q_0^{\dot\alpha},H_0]
=\int_0{dw\over 2\pi}\int_w{dz\over 2\pi}\sqrt{iz}(-w)\biggl(
-3{\gamma^i_{\alpha\dot\alpha}\theta^\alpha\partial X^i(w)
\over(z-w)^2}
-{\partial(\gamma^i_{\alpha\dot\alpha}\theta^\alpha\partial X^i)(w)
\over z-w}\biggr).}
After expanding $\sqrt{z}$ and picking up the residue, we have to
integrate by parts in the second term to show both terms cancel
identically.

Relations for the other supercharge $\tilde Q_0^{\dot\alpha}$ can be
found in the same way.
Since holomorphic fields and anti-holomorphic fields do not have
non-trivial OPE, we also find our choice for two supercharges is
good: they anti-commute with each other,
$\{Q_0^{\dot\alpha},\tilde Q_0^{\dot\beta}\}=0$.

Now let us proceed to the first non-trivial order.
For this purpose, let us rewrite \Hone\ and \Qone\ as
\eqn\oneth{\eqalign{
&H_1=\int{dz\over 2\pi}
\sqrt{iz}\overline{iz}^{3/2}
([\tau^i\bar\tau^j]\Sigma^i\bar\Sigma^j),\cr
&Q_1^{\dot\alpha}=-\sqrt{2}\int{dz\over 2\pi}\overline{iz}^{3/2}
([\sigma\bar\tau^j]\Sigma^{\dot\alpha}\bar\Sigma^j).}}
Here we drop the string bit indices $(mn)$ again.
We shall fix one particular pair $(mn)$ of the string bits in the
first order expression \oneth\ and rewrite the quadratic terms of this
pair in \zerohol\ into sum and difference:
\eqn\rewrite{{\cal O}_m{\cal P}_m+{\cal O}_n{\cal P}_n
={{\cal O}_m+{\cal O}_n\over\sqrt{2}}
{{\cal P}_m+{\cal P}_n\over\sqrt{2}}
+{{\cal O}_m-{\cal O}_n\over\sqrt{2}}
{{\cal P}_m-{\cal P}_n\over\sqrt{2}},}
with ${\cal O}$ and ${\cal P}$ being $X^i$ or $\theta^\alpha$. 
Since the first order expression \oneth\ with indices $(mn)$ exchanges
two string bits $m$ and $n$, it have non-trivial OPE only with the
difference term in \rewrite.
After this calculation we sum up over our choice $(mn)$.
In other words, our expression at the zeroth order \zerohol\ has the
implicit indices $(mn)$ now, meaning each operator is the difference
of that with index $m$ and that with $n$.

Now we need some non-trivial OPE.
These OPE are evaluated in our appendices.
\eqn\someOPE{\eqalign{
\partial X^i(z)\cdot\sigma(0)&\sim{\eta^*\over\sqrt{z}}
\Bigl(\tau^i(0)+O(z)\Bigr),\cr
\partial X^i(z)\cdot\tau^j(0)
&\sim{\eta^*\over\sqrt{z}}\biggl({\delta^{ij}\over 2z}\sigma(0)
+{\delta^{ij}\over 4}\partial\sigma(0)+%{1\over\sqrt{2}}
\Bigl(\tau^{ij}-{\delta^{ij}\over 4}\partial\sigma\Bigr)(0)
+O(z)\biggr),\cr
%\partial X^i(z)\cdot\bar\tau^j(0,0)
%&\sim{\eta^*\over\sqrt{z}}\Bigl(\chi^{ij}(0,0)+O(z)\Bigr),\cr
\theta^\alpha(z)\cdot\Sigma^{\dot\alpha}(0)
&\sim{\eta^*\over\sqrt{2z}}
\Bigl(\gamma^i_{\alpha\dot\alpha}\Sigma^i(0)+O(z)\Bigr),\cr
\theta^\alpha(z)\cdot\Sigma^i(0)
&\sim{\eta^*\over\sqrt{2z}}
\biggl(\gamma^i_{\alpha\dot\alpha}\Sigma^{\dot\alpha}(0)
+{z\over 2}\gamma^i_{\alpha\dot\alpha}\partial\Sigma^{\dot\alpha}(0)
+\langle?\rangle zu^{i\alpha}_{\dot\alpha\dot\beta\dot\gamma}
\Sigma^{\dot\alpha}\Sigma^{\dot\beta}\Sigma^{\dot\gamma}(0)
+O(z^2)\biggr),}}
with $\eta^*=\exp(-i\pi/4)$ and
$u^{i\alpha}_{\dot\alpha\dot\beta\dot\gamma}
=-\gamma^{ij}_{[\dot\alpha\dot\beta}\gamma^j_{\dot\gamma]\alpha}$.
As we noted previously, the twist fields do not decouple between the
holomorphic parts and the anti-holomorphic parts in general.
We write down the first two formulas in this way because they hold
independently of excitations by anti-holomorphic bosons.
Also, $\langle?\rangle$ in the third relation means a coefficient
which is not fixed.
Fortunately, this is not necessary in our calculation.
The doubly excited twist field $\tau^{ij}$ relates to
$\alpha_{-1/2}^i\alpha_{-1/2}^j|0\rangle_{\rm AP}$ and the expression
in the parenthesis $(\tau^{ij}-\delta^{ij}\partial\sigma/4)$ denotes
the traceless combination, which has no singularity in OPE with
$\sigma$ itself.
Also, the twist field and the spin field are primary fields, so we
have
\eqn\primaryfield{\eqalign{
T_b(z)\cdot\sigma(0)
&\sim{1\over 2z^2}\sigma(0)
+{1\over z}\partial\sigma(0)+O(z^0),\cr
T_f(z)\cdot\Sigma^{\dot\alpha}(0)
&\sim{1\over 2z^2}\Sigma^{\dot\alpha}(0)
+{1\over z}\partial\Sigma^{\dot\alpha}(0)+O(z^0),}}
with $T_b$ and $T_f$ being the boson part and fermion
part of the energy momentum tensor, respectively:
\eqn\enemom{T_b=-{1\over 2}:\partial X^i\partial X^i:,\qquad
T_f=-{1\over 2}:\theta^\alpha\partial\theta^\alpha:.}

Using all these OPE, a proof of supersymmetry algebra is possible.
With the help of following two formulas,
\eqn\calc{\eqalign{
[Q_0^{\dot\alpha},\sigma\Sigma^{\dot\beta}(w)]
&=-{\sqrt{iw}\over\sqrt{2}}
\gamma^i_{\alpha\dot\alpha}\gamma^j_{\alpha\dot\beta}
\tau^i\Sigma^j(w),\cr
\{Q_0^{\dot\alpha},\tau^i\Sigma^i(w)\}
&=-{\sqrt{2i}\over\sqrt{w}}\Bigl(\sigma\Sigma^{\dot\alpha}(w)
+w\partial(\sigma\Sigma^{\dot\alpha})(w)\Bigr),}}
we can prove the first non-trivial order of the supersymmetry algebra,
the second line of \susyexpand,
\eqn\secondline{
\{Q^{\dot\alpha}_0,Q^{\dot\beta}_1\}
+\{Q^{\dot\alpha}_1,Q^{\dot\beta}_0\}
=2H_1\delta^{\dot a\dot b},\qquad
[Q^{\dot\alpha}_0,H_1]+[Q^{\dot\alpha}_1,H_0]=0,}
as well as independence between two supercharges:
\eqn\independ{\{Q_0^{\dot\alpha},\tilde Q_1^{\dot\beta}\}
+\{Q_1^{\dot\alpha},\tilde Q_0^{\dot\beta}\}=0.}

\newsec{Discussion on contact terms}
Now we have proved the supersymmetry algebra at the first non-trivial
order.
Let us proceed to discuss the contact terms in the systematic method
as summarized in section 2.

As is obvious by now, our strategy to the contact terms is to
calculate $\{Q^{\dot\alpha}_1,Q^{\dot\beta}_1\}$ in \susyexpand\ and
try to determine the next order $Q_2^{\dot\alpha}$ and $H_2$.
Here we have not succeeded in this goal.
We only discuss some problems of this calculation.

We have three types of terms for
$\{Q^{\dot\alpha}_1,Q^{\dot\beta}_1\}$, depending on relative
interaction points:
$\{Q^{\dot\alpha}_{1(kl)},Q^{\dot\beta}_{1(kl)}\}$,
$\{Q^{\dot\alpha}_{1(kl)},Q^{\dot\beta}_{1(lm)}\}$, and
$\{Q^{\dot\alpha}_{1(kl)},Q^{\dot\beta}_{1(mn)}\}$.
As discussed in \DijkgraafNW, on the SYM side of the PP-wave/SYM
correspondence the Feynman diagram looks like the second type.
This leads them to make a proposal for the contact term and
investigate the $Z_3$ twist field carefully.
However, in order to match with the SYM calculation, on the
light-cone superstring field theory side \RoibanXR\GomisKJ\ we also
have to consider the contribution somewhat similar to the first type.
This type of terms corresponds to the interaction where one string
splits into two and these two strings joins into one again at the same
interaction point.
The third type is trivial.
Since two operators in the OPE of $Q_1$ relate to different string
bits, there is no singularity appearing in the OPE.
Therefore, this type of contact term should be vanishing.

Let us concentrate on the first type
$\{Q^{\dot\alpha}_{1(kl)},Q^{\dot\beta}_{1(kl)}\}$.
Note that since our Hamiltonian and supercharges, \Hone\ and \Qone,
are not written in the expression of conserved charges \charge, we
cannot deform the integral contour arbitrarily.
However, we are considering the commutation relation at the same
radius.
Hence, apart from the contribution where two operators are close to
each other, two integral contours in the commutation relation should
cancel each other.
Therefore, we are allowed to consider the OPE of these operators.
The leading OPE are read off from the dimensions:
\eqn\result{\eqalign{
[\sigma\bar\tau^i](z,\bar z)\cdot[\sigma\bar\tau^j](0,0)
&\sim{\delta^{ij}\over 2z\bar z^2}+O(z^0),\cr
\Sigma^{\dot\alpha}(z)\cdot\Sigma^{\dot\beta}(0)
&\sim{\delta^{\dot\alpha\dot\beta}\over z}+O(z^0),\cr
\bar\Sigma^i(\bar z)\cdot\bar\Sigma^j(0)
&\sim{\delta^{ij}\over\bar z}+O(\bar z^0).}}
The subleading orders are also relevant after multiplying the twist
fields and the spin fields.

We would like to integrate these OPE over $z$ and $w$ and find exactly
the commutator $\{Q^{\dot\alpha}_{1(kl)},Q^{\dot\beta}_{1(kl)}\}$.
However, there are some problems.
First of all, as we have noted, we cannot deform the contour
arbitrarily. 
Therefore, the commutators at the same radius will depend on the
regularization.
At present we do not have a good principle to choose the
regularization.
A natural one may be to encircle the singularity after cancelling the
contribution where two operators are separated far away, as the usual
contour deformation technique.
If we adopt this regularization, the second problem appears.
The integration of the most singular term
\eqn\singular{\int_0{dw\over 2\pi i}\int_w{dz\over 2\pi i}
\bar z^{3/2}\bar w^{3/2}{1\over(z-w)^2(\bar z-\bar w)^3},}
gives divergent contribution.
We need an interpretation for it.

This divergence in $\{(Q_1)_{(kl)},(Q_1)_{(kl)}\}$ may not be
surprising.
We can perform a similar calculation in the PP-wave string field
theory.
When we match two geometrical interaction vertices $|V\rangle$, we
need the formula:
\eqn\formula{\exp\biggl({1\over 2}aNa\biggr)
\exp\biggl({1\over 2}a^\dagger Ma^\dagger\biggr)|0\rangle
={1\over\sqrt{\det(1-MN)}}\exp
\biggl({1\over 2}a^\dagger M{1\over 1-NM}a^\dagger\biggr)|0\rangle.}
However, if we take the mass parameter $\mu$ of the PP-wave background
to infinity, these matrices $M, N$ are the same orthonormal matrices.
Therefore, it looks singular in this limit.

The situation for the second type $\{(Q_1)_{(kl)},(Q_1)_{(lm)}\}$ is
more complicated.
Even thought it is easy to guess the OPE between the twist fields
$[\sigma\bar\sigma]_{(kl)}$ and $[\sigma\bar\sigma]_{(lm)}$ from the
multiplication of the permutation group $(kl)(lm)=(klm)$ and the
dimension counting, it is not so easy to calculate the OPE of the
excited twist fields $[\sigma\bar\tau^i]_{(kl)}(z,\bar z)
\cdot[\sigma\bar\tau^j]_{(lm)}(0,0)$ because they are excited by
different fields.

\bigskip\bigskip
\centerline{{\bf Acknowledgements}}\nobreak
We would like to thank Jaume Gomis, Lubos Motl, Hirosi Ooguri, John
Schwarz and especially Takuya Okuda for useful discussions.
We are also grateful to Yuji Okawa for pointing out a mistake in the
previous version of this paper.
This research was supported in part by the DOE grant 
DE-FG03-92-ER40701.

\bigskip\bigskip
\appendix{A}{OPE of twist fields}
In this appendix, we shall show the first line of \someOPE.
We begin with showing
\eqn\dxsbs{
\partial X^i(z)\cdot[\sigma\bar\sigma](0,0)
\sim{\eta^*\over\sqrt{z}}\Bigl([\tau^i\bar\sigma](0,0)+O(z)\Bigr),}
with no excitations by the anti-holomorphic bosons.
For this purpose let us consider the four point function
$\langle[\sigma\bar\sigma](z_1,\bar z_1)\partial X^i(z_2)
\partial X^j(z_3)[\sigma\bar\sigma](z_4,\bar z_4)\rangle$.
By taking two bosons $\partial X^i(z)$ close to two twist fields
$[\sigma\bar\sigma](z,\bar z)$, respectively, we can read off the
above OPE.

The calculation of the four point functions
$\langle[\sigma\bar\sigma](z_1,\bar z_1)\partial X^i(z_2)
\partial X^j(z_3)[\sigma\bar\sigma](z_4,\bar z_4)\rangle$ is standard
and can be found, for example in the textbook \DiFrancescoNK.
Since it will also be necessary in the next appendix we shall
recapitulate it briefly in the following.

Using the OPE of bosons \free\ and the mode expansion of bosons in the
twisted sector
\eqn\bosmode{i\partial X^i(z)
=\sum_{r\in Z+1/2}{\alpha_r\over z^{r+1}},}
we find the commutation relation is unchanged as in the untwisted
sector
\eqn\boscomm{[\alpha_r^i,\alpha_s^j]=r\delta^{ij}\delta_{r+s},}
except that the indices run over half-integers now.
Hence, we can compute the two point function in the twisted sector
\eqn\sxxslim{{}_{\rm AP}\langle 0|\partial X^i(z_2)
\partial X^j(z_3)|0\rangle_{\rm AP}
=-{\sqrt{z_2/z_3}+\sqrt{z_3/z_2}\over 2(z_2-z_3)^2}\delta^{ij},}
where $|0\rangle_{\rm AP}$ denotes the anti-periodic twisted ground
state $[\sigma\bar\sigma](0,0)|0\rangle$.
In other words, this quantity can be interpreted as the four point
function,
\eqn\interpretfour{\lim_{z_1\to\infty,z_4\to 0}z_1\bar z_1
\langle[\sigma\bar\sigma](z_1,\bar z_1)\partial X^i(z_2)
\partial X^j(z_3)[\sigma\bar\sigma](z_4,\bar z_4)\rangle.}
Restoring the dependence on $z_1$ and $z_4$ from the general argument
of four point functions in two dimensional CFT, we can rewrite the
previous results in a more democratic way:
\eqn\sxxs{\langle[\sigma\bar\sigma](z_1,\bar z_1)\partial X^i(z_2)
\partial X^j(z_3)[\sigma\bar\sigma](z_4,\bar z_4)\rangle
=-{1\over z_{14}z_{23}^2\bar z_{14}}
\sqrt{z_{13}z_{24}\over z_{12}z_{34}}
\biggl(1-{z_{14}z_{23}\over 2z_{13}z_{24}}\biggr)\delta^{ij},}
with $z_{ij}\equiv z_i-z_j$.

Now let us take the limit $|z_{12}|, |z_{34}|\ll |z_{14}|$ in \sxxs.
\eqn\sxxsother{
\langle[\sigma\bar\sigma](z_1,\bar z_1)\partial X^i(z_2)
\partial X^j(z_3)[\sigma\bar\sigma](z_4,\bar z_4)\rangle
\simeq-{\delta^{ij}\over 2z_{14}^2\bar z_{14}\sqrt{z_{12}z_{34}}}.}
This means that if we choose the excited twist field 
$[\tau^i\bar\sigma](z,\bar z)$ to correspond to the state
$\alpha_{-1/2}^i|0\rangle_{\rm AP}$, then we have the OPE \dxsbs\ and
\eqn\taudef{
[\tau^i\bar\sigma](z,\bar z)\cdot[\tau^j\bar\sigma](0,0)
\sim{\delta^{ij}\over 2z^2\bar z}+O(z^{-1}).}

As is obvious by repeating the same calculation, this calculation does
not depend on the excitations by anti-holomorphic bosons
$\bar\alpha_r^i$.
Therefore, we can rewrite the previous result symbolically as
\eqn\dXs{\partial X^i(z)\cdot\sigma(0)
\sim{\eta^*\over\sqrt{z}}\Bigl(\tau^i(0)+O(z)\Bigr).}
The first line in \someOPE\ is shown in this way.

\appendix{B}{OPE of excited twist fields}
This appendix is devoted to the second line of \someOPE.
As in the previous appendix, this calculation also does not depend on
the excitations by anti-holomorphic bosons $\bar\alpha_r^i$.
Therefore we drop the anti-holomorphic dependence from the beginning.
Note again that this is purely symbolic notation.
In other computations the holomorphic part and the anti-holomorphic
part do not decouple in general.

Since the excited twist field $\tau^i(z)$ corresponds to
$\alpha^i_{-1/2}|0\rangle_{\rm AP}$, the four point function
\eqn\excited{
\lim_{z_1\to\infty,z_4\to 0}z_1^2
\langle\tau^i(z_1)\partial X^k(z_2)
\partial X^l(z_3)\tau^j(z_4)\rangle,}
can be calculated as
\eqn\excitedfour{\eqalign{
&{}_{\rm AP}\langle 0|\alpha_{1/2}^i\partial X^k(z_2)
\partial X^l(z_3)\alpha_{-1/2}^j|0\rangle_{\rm AP}\cr&\qquad=
-{1\over 4}{\sqrt{z_2/z_3}+\sqrt{z_3/z_2}\over(z_2-z_3)^2}
\delta^{ij}\delta^{kl}
-{1\over 4}{\sqrt{z_2/z_3}\over z_2z_3}\delta^{ik}\delta^{jl}
-{1\over 4}{\sqrt{z_3/z_2}\over z_2z_3}\delta^{il}\delta^{jk}.}}
Again from the general arguments of four point functions, we can
rewrite it into
\eqn\excitedfourpt{\eqalign{
&\langle\tau^i(z_1)\partial X^k(z_2)
\partial X^l(z_3)\tau^j(z_4)\rangle\cr
&\qquad=-{1\over 4}\sqrt{{z_{13}z_{24}\over z_{12}z_{34}}}
\biggl\{\biggl(1+{z_{12}z_{34}\over z_{13}z_{24}}\biggr)
{\delta^{ij}\delta^{kl}\over(z_{14}z_{23})^2}
+{\delta^{ik}\delta^{jl}\over z_{12}z_{34}z_{13}z_{24}}
+{\delta^{il}\delta^{jk}\over(z_{13}z_{24})^2}\biggr\}.}}

As in the previous appendix, we consider the OPE of $\partial X^i$ and
$\tau^j$.
Note that, from the Lorentz symmetry and dimension analysis, terms
that we can have in the operator expansion should be restricted to
$\delta^{ij}\sigma/z^{3/2}$, $\delta^{ij}\partial\sigma/z^{1/2}$ and
$(\tau^{ij}-\partial\sigma/4)/z^{1/2}$ up to $O(z^{1/2})$.
Here $\tau^{ij}$ corresponds to
$\alpha_{-1/2}^i\alpha_{-1/2}^j|0\rangle_{\rm AP}$ and the
subtraction in the last term comes from the traceless condition.
Note that the derivative $\partial\sigma$ is the descendent of the
twist field $L_{-1}\sigma$, hence this field corresponds to 
$(1/2)\alpha_{-1/2}^i\alpha_{-1/2}^i|0\rangle_{\rm AP}$.
These correspondence implies the OPE
\eqn\tauij{\eqalign{\partial\sigma(z)\cdot
\Bigl(\tau^{ij}-{\delta^{ij}\over 4}\partial\sigma\Bigr)(0)
&\sim 0+O(z^{-4}),\cr
\Bigl(\tau^{ik}-{\delta^{ik}\over 4}\partial\sigma\Bigr)(z)
\cdot\Bigl(\tau^{jl}-{\delta^{jl}\over 4}\partial\sigma\Bigr)(0)
&\sim{\delta^{ik}\delta^{jl}
-4(\delta^{ij}\delta^{kl}+\delta^{il}\delta^{jk})\over 16z^3}
+O(z^{-4}).}}

In order to determine the coefficients, let us take the limit
$|z_{12}|, |z_{34}|\ll |z_{14}|$ again in \excitedfourpt:
\eqn\excitedother{\eqalign{&\langle\tau^i(z_1)
\partial X^k(z_2)\partial X^l(z_3)\tau^j(z_4)\rangle
\simeq-{1\over 4z_{14}(z_{12}z_{34})^{3/2}}\cr
&\qquad\times\biggl(
\delta^{ik}\delta^{jl}\Bigl(1+{1\over 2}{z_{12}\over z_{14}}
+{1\over 2}{z_{34}\over z_{14}}
+{1\over 4}{z_{12}z_{34}\over z_{14}^2}\Bigr)
+\delta^{ij}\delta^{kl}{z_{12}z_{34}\over z_{14}^2}
+\delta^{il}\delta^{jk}{z_{12}z_{34}\over z_{14}^2}\biggr),}}
and read off the coefficients from this expansion.
The result is
\eqn\dxtau{\partial X^i(z)\cdot\tau^j(0)
\sim{\eta^*\over\sqrt{z}}\biggl({\delta^{ij}\over 2z}\sigma(0)
+{\delta^{ij}\over 4}\partial\sigma(0)
+\Bigl(\tau^{ij}-{\delta^{ij}\over 4}\partial\sigma\Bigr)(0)
+O(z)\biggr).}

\appendix{C}{OPE of spin fields}
In this appendix, we shall calculate the OPE of spin fields, the last
one in \someOPE,
\eqn\tS{\theta^\alpha(z)\cdot\Sigma^i(-z)\sim{\eta^*\over\sqrt{4z}}
\Bigl(\gamma^i_{\alpha\dot\alpha}\Sigma^{\dot\alpha}(0)
+\langle?\rangle zu^{i\alpha}_{\dot\alpha\dot\beta\dot\gamma}
:\Sigma^{\dot\alpha}\Sigma^{\dot\beta}\Sigma^{\dot\gamma}:(0)
+O(z^2)\Bigr).}
The method adopted in this appendix is completely different from the
previous two appendices.
Since we can represent the spin fields by free bosons, our strategy is
first to restrict terms which can appear in the expansion and then
identify the coefficients by calculating the OPE of some typical terms
with the free field realization.

Let us consider
$\gamma^i_{\alpha\dot\alpha}\theta^\alpha(z)\cdot\Sigma^i(-z)$
for the evaluation of \tS.
Since only terms having the same Lorentz indices and correct conformal
dimensions can appear in the expansion, the possible terms up to
$O(z^{3/2})$ are restricted to
\eqn\guess{{1\over\sqrt{z}}\Sigma^{\dot\alpha},\qquad
\sqrt{z}\partial\Sigma^{\dot\alpha},\qquad
\sqrt{z}t^{ii}_{\dot\alpha\dot\beta\dot\gamma\dot\delta}
\Sigma^{\dot\beta}\Sigma^{\dot\gamma}\Sigma^{\dot\delta},}
with $t^{ij}_{\dot\alpha\dot\beta\dot\gamma\dot\delta}$ defined by
\eqn\tij{t^{ij}_{\dot\alpha\dot\beta\dot\gamma\dot\delta}
=\gamma^{ik}_{[\dot\alpha\dot\beta}\gamma^{jk}_{\dot\gamma\dot\delta]}.}
Since $t^{ij}_{\dot\alpha\dot\beta\dot\gamma\dot\delta}$ is traceless,
the last term vanishes identically.
%It is well-known that in the product of two ${\bf 8_s}$
%representation ${\bf 8_s \otimes 8_s=1+28+35}$, the representation
%${\bf 35}$ can be written as self-dual anti-symmetric tensor.
%Here $t^{ij}_{\alpha\beta\gamma\delta}$ is dual form of this
%relation: it relates two vector indices to four completely
%anti-symmetric spinor indices.

We have only to determine the coefficient of each term by the free
field realization.
If we combine the real fermions into complex ones as
\eqn\bosonize{\eqalign{
\Theta^A
\equiv{1\over\sqrt{2}}(\theta^A+i\theta^{A+4}),&\qquad
\Theta^{\bar A}
\equiv{1\over\sqrt{2}}(\theta^A-i\theta^{A+4}),\cr
{\cal S}^A
\equiv{1\over\sqrt{2}}(\Sigma^{2A-1}+i\Sigma^{2A}),&\qquad
{\cal S}^{\bar A}
\equiv{1\over\sqrt{2}}(\Sigma^{2A-1}-i\Sigma^{2A}),\cr
{\cal S}^{\dot A}
\equiv{1\over\sqrt{2}}(\Sigma^{\dot A}+i\Sigma^{\dot A+4}),&\qquad
{\cal S}^{\dot{\bar A}}
\equiv{1\over\sqrt{2}}(\Sigma^{\dot A}-i\Sigma^{\dot A+4}),}}
these complex fermions are realized in terms of free bosons $\phi^A
(A=1,\ldots,4)$ by,
\eqn\bosonization{
\Theta^A=e^{iq^A_B\phi^B},\;\Theta^{\bar A}=e^{-iq^A_B\phi^B},\;
{\cal S}^A=e^{i\phi^A},\;{\cal S}^{\bar A}=e^{-i\phi^A},\;
{\cal S}^{\dot A}=e^{iq^{\dot A}_B\phi^B},\;
{\cal S}^{\dot{\bar A}}=e^{-iq^{\dot A}_B\phi^B},}
with
\eqn\qs{\eqalign{
&q^1={1\over 2}(1,1,1,1),\quad
q^2={1\over 2}(1,1,-1,-1),\quad
q^3={1\over 2}(1,-1,1,-1),\quad
q^4={1\over 2}(1,-1,-1,1),\cr
&q^{\dot 1}={1\over 2}(-1,1,1,1),\quad
q^{\dot 2}={1\over 2}(1,-1,1,1),\quad
q^{\dot 3}={1\over 2}(1,1,-1,1),\quad
q^{\dot 4}={1\over 2}(1,1,1,-1).}}

All the coefficients are determined by considering various examples of
the OPE with the free field realization.
For example, to have $e^{iq^{\dot 1}_B\phi^B}$ we need to consider,
\eqn\dvanish{\eqalign{
e^{iq^1_B\phi^B}(z)\cdot e^{-i\phi^1}(-z)&\sim e^{iq^{\dot 1}_B\phi^B}
\Bigl(1+iz\partial(q^1_B\phi^B+\phi^1)(0)+O(z^2)\Bigr),\cr
e^{-iq^2_B\phi^B}(z)\cdot e^{i\phi^2}(-z)&\sim e^{iq^{\dot 1}_B\phi^B}
\Bigl(1+iz\partial(-q^2_B\phi^B-\phi^2)(0)+O(z^2)\Bigr),\cr
e^{-iq^3_B\phi^B}(z)\cdot e^{i\phi^3}(-z)&\sim e^{iq^{\dot 1}_B\phi^B}
\Bigl(1+iz\partial(-q^3_B\phi^B-\phi^3)(0)+O(z^2)\Bigr),\cr
e^{-iq^4_B\phi^B}(z)\cdot e^{i\phi^4}(-z)&\sim e^{iq^{\dot 1}_B\phi^B}
\Bigl(1+iz\partial(-q^4_B\phi^B-\phi^4)(0)+O(z^2)\Bigr).}}
Summing over the RHS, we find the coefficient of
$z\partial\Sigma^{\dot\alpha}$ also vanishes.
Our goal of this appendix \tS\ follows in this way.
\listrefs
\end